# A unified picture of roto-translational dynamics in aqueous polyatomic ions

Puja Banerjee and Biman Bagchi*

Solid State and Structural Chemistry Unit, Indian Institute of Science, Bangalore-560012, India

## Abstract

*Mode-coupling theory provides a unified description of the rotational and translational dynamics of polyatomic ions. These molecular ions are distinct from usual models of ion diffusion, such as $K^+$, $Cl^-$ etc., and also different from rotational dynamics of dipolar molecules often modeled in dielectric continuum models as point dipoles. Both these approaches are untenable for polyatomic ions. Here rotational and translational dynamics are so strongly coupled that one obtains a more coherent description by treating them together. We carry out theoretical and computational studies of a series of well-known polyatomic ions, namely sulfate, nitrate and acetate ions. All the three ions exhibit different rotational diffusivity, with that of nitrate ion being considerably larger than the other two. They all defy the hydrodynamic laws of size dependence. Study of the local structure around the ions provides valuable insight into the origin of these differences. We carry out a detailed study of the rotational diffusion of these ions by extensive computer simulation and using the theoretical approaches of the dielectric friction developed by Fatuzzo-Mason (FM) and Nee-Zwanzig (NZ), and subsequently generalized by Alavi and Waldeck. We develop a self-consistent mode-coupling theory (SC-MCT) formalism that helps elucidating the role of coupling between translational and rotational motion of these ions. In fact, these two motions self-consistently determine the value of each other. The RISM-based MCT suggests an interesting relation between the torque-torque and the force-force time correlation function with the proportionality constant being determined by the geometry and the charge distribution of the polyatomic molecule. We point out several parallelism between the theories of translational and rotation friction calculations of ions in dipolar liquids.*

Corresponding author. Email: bbagchi@sscu.iisc.ernet.in

# I. Introduction

Ions in aqueous solution are omnipresent in nature. Oceans are vast electrolyte solutions, our human body contains many ions that play crucial role in life processes, ions are also important in chemical industry. Naturally the study of ions in dipolar liquids have been central to physical chemistry and chemical physics and have continually drawn interest. However, attention has remained focussed for a long time on monatomic ions like rigid alkali cations and monovalent haliode ions [1-3,4,5]. The study of polyatomic ions have been neglected.

However, polyatomic ions play important role both in chemistry and biology. They are common electrolytes, they form the buffer solutions in chemistry and biology and they are extremely important in chemical industry. For example, the economic health of a nation is gauged by the amount of sulfuric acid it produces per year. Interestingly, many of the important polyatomic ions are anions, like sulfate, phosphate, nitrate and acetate.

Unlike the rigid spherical ions often studied theoretically, polyatomic ions are unique in the sense that they can rotate. In addition, they contain charges distributed over constituent atoms. Some of these atoms are often close to the surface of the molecule and therefore interact with the solvent molecules quite strongly. Structure, dynamics and transport properties of these anions offer both interesting and important properties that have not been explored theoretically at a quantitative level, although substantial amount of experimental results exist. In this work, we shall carry out both computer simulations and analytical theoretical studies of three important anions in water.

Transport properties of polyatomic anions are different from the monatomic anions, like fluoride, chloride, bromide or the rigid alkali cations like potassium, sodium etc. The polyatomic ions are usually larger than the monatomic ions, non-spherical in shape, contain distributed charges, with some of the charges placed close to the surface. This allows stronger interaction with charges of water molecules. In addition, these molecules execute both rotational and translational motions which are obviously coupled.

As mentioned above, the dynamics of polyatomic ions (both rotational and translational dynamics) of polyatomic ions have hardly been studied theoretically, although a considerable amount of experimental results exist. The dynamical studies of these ions offer interesting challenges.
To be precise, the objectives of the present study are the following:

(i) We carry out detailed MD simulation to analyse the rotational diffusion of polyatomic ions in water.

(ii) Given the non-spherical shape and distributed charges, we want to develop a theory for their rotational diffusion. To begin with, we apply a continuum model of rotational dielectric friction first developed by Fatuzzo and Mason[6], and by Nee and Zwanzig[7], and later greatly generalized by Alavi and Waldeck[8].

(iii) We present a detailed study of the rotational dynamics of sulphate ion for the first time.

(iv) Computer simulations clearly show the translation-rotational coupling and microscopic structural features that determines their dynamics.

(v) We develop a mode coupling theory formalism to describe the effects of translational-rotational coupling on the rotational dielectric friction.

The theoretical study of rotational friction has a long history [9-13]. It started with the simplest model of hydrodynamics, Debye-Stokes-Einstein (DSE) model where friction depends only on the size (ionic radius, R) and shape of the solute and viscosity of the bulk liquid ($\eta$).

$$\zeta_{DSE} = 8\pi\eta R^3 \qquad (1)$$

However, the neglect of molecular level interaction between solute and solvent often leads to the breakdown of this model. One example of this kind is the charged (or, dipolar) solute in dipolar solvent where the motion of the solute can be coupled to the dielectric response of the medium. But, simple hydrodynamic theories treat the rotational dynamics of polar and nonpolar solutes equally. However, the additional friction arising in case of polar medium for charged solute is known as "dielectric friction". This is closely connected to the solvation dynamics of the polar medium [13-15].

The main motivation for the study of rotational dielectric friction comes from non-exponential dielectric relaxation of dipolar liquids. However, the systems we study here are molecular ions, and they have no counter-part either in dipolar liquids or in monatomic rigid ions. In an recent experiment, the breakdown of Debye-Stokes-Einstein model has been demonstrated for guanidine hydrochloride in water and mixtures of carbon disulfide with hexadecane by using optical Kerr-effect spectroscopy [16]. However, as mentioned earlier, these systems have not been studied properly by theory and computation. One should be able to combine certain aspects from earlier theoretical developments. To this goal, we briefly review the theoretical developments both in the area of rotational dielectric friction on rotating dipoles (**Figure 1**) and translational dielectric friction on ions, both in liquids.

In a famous work, Hu and Zwanzig calculated frictional drag on a rotating ellipsoid (prolate and oblate) by hydrodynamic calculation [17] and they showed that hydrodynamics can be used even at a molecular level. This model has been extensively used by many studies to explain experimental results of rotational relaxation [18]. The pioneering work on rotational dieletric friction that shaped all subsequent discussions were made by Fatuzzo and Mason [6] and Nee and Zwanzig [7]**.** These groups started with a decomposition of the total rotational friction in terms of a hydrodynamic term due to viscosity and a term due to polarization fluctuations

$$\zeta_{Rot} = \zeta_{DSE} + \zeta_{R,DF} \qquad (2)$$

where $\zeta_{DSE}$ is the hydrodynamic contribution to rotational friction obtained from Debye-Stokes-Einstein(DSE) relation (Eq. (1)) and $\zeta_{R,DF}$ is the rotational dielectric friction.

Nee and Zwanzig developed an expression for dielectric friction on a point dipole rotating at the centre of a spherical cavity in a frequency dependent dielectric medium[7].

$$\zeta_{DF}(\omega) = \frac{2k_B T}{i\omega} \frac{(\varepsilon_0 - \varepsilon_\infty)}{\varepsilon_0} \frac{(\varepsilon(\omega) - \varepsilon_0)}{(2\varepsilon(\omega) + \varepsilon_\infty)} \quad (3)$$

This expression can be generalised with explicit dependence on cavity size, "a" and molecular dipole moment, μ by using Onsager's expression of static dielectric constant[9]

$$\zeta_{DF}(\omega) = \frac{6\mu^2}{i\omega a^3} \left\{ \frac{\varepsilon(\omega) - \varepsilon_0}{(2\varepsilon_0 + 1)[2\varepsilon(\omega) + 1]} \right\} \quad (4)$$

The resulting expression in the limit of slow solute rotation or, zero frequency limit is given by

$$\zeta_{NZ} = \frac{6\mu^2 \tau_D (\varepsilon_0 - \varepsilon_\infty)}{a^3 (2\varepsilon_0 + \varepsilon_\infty)^2} \quad (5)$$

where μ is the solute dipole moment, "a" is the cavity radius, $\tau_D$ is the Debye relaxation time and $\varepsilon_s$ and $\varepsilon_\infty$ are the static dielectric constant and the high-frequency dielectric constant respectively.

In an detailed study, Zhou ad Bagchi (ZB) investigated the validity of continuum model by using computer simulation of a brownian dipolar lattice introduced earlier by Zwanzig[19][20][21]. ZB pointed out that the derivation of NZ expression involves three approximations: 1) the generalised diffusion equation using the density functional theory description for a single particle orientation, 2) the relation between single particle rotational dynamics and dielectric function through the effective medium description and 3) the relation between dielectric function and frequency dependent dielectric constant. It was found that NZ works quite well for single particle dynamics unlike collective dynamics but the continuum model description for torque-torque correlation decays much faster. Alavi and Waldeck [8][22][23][24] extended the Nee-Zwanzig theory and calculated dielectric friction for an arbitrary charge distribution in a spherical cavity that will be discussed later in detail.

Earlier, in a series of works, van der Zwan and Hynes calculated wavevector and frequency dependent dielectric function of a polar liquid by using a semiempirical method[25, 26]. In this method, the difference between Stokes shift, S (measures the magnitude of polarisation field induced by solvent) and the solvation time, $\tau_s$ (measures the lag of this polarisation field) is used to obtain the dielectric friction. Another important contribution to this field came from Evans who used Enskog kinetic theory description to measure rotational and translational friction on non-dipolar spherical particles[27]. In addition, Chandler developed a theory for the rotational diffusion based on a model system of rough hard spheres interacting through binary collisions [28]. But this model is limited for a system of only spherical particles having instantaneous collisions.

An extended molecular theory of orientational relaxation of solvent around a spherical solute has been developed by our group [29][30]. The theory is based on the nonlinear Smoluchowski equation for rotational and translational diffusion and can be written as

$$\frac{\partial}{\partial t}\delta\rho(r,\Omega,t) = D_R \nabla_\Omega^2 \delta\rho(r,\Omega,t) - D_T \nabla_T^2 \delta\rho(r,\Omega,t) - D_R (\rho_0/4\pi) \nabla_\Omega^2 \int dr'd\Omega' c(r-r',\Omega,\Omega')\delta\rho(r',\Omega',t)$$
$$+ D_T (\rho_0/4\pi) \nabla_T^2 \int dr'd\Omega' c(r-r',\Omega,\Omega')\delta\rho(r',\Omega',t)$$

(6)

where $D_T$ and $D_R$ are the translational and rotational diffusivity, respectively and $\nabla_T$ and $\nabla_\Omega$ are the spatial and rotational gradient operator. In the limit of slow solute reorientation (when solute rotation is slower than the solvent dynamics), dielectric friction for rotation experienced by the solute can be expressed as a time integral of torque-torque time correlation function (TTTCF)

$$\zeta_{R,DF} = \frac{1}{2k_B T} \int_0^\infty dt \langle N(\Omega,0) \cdot N(\Omega,t) \rangle \qquad (7)$$

where $N(\Omega,t)$ is the total torque experienced by the solute molecule with orientation $\Omega$ at time t. Earlier we showed that, using classical density functional theory, the expression for frequency dependent dielectric friction on solute due to rotational motion can be obtained from TTTCF[31,32]

$$\zeta_{DF}(\omega) = \frac{k_B T}{2(2\pi)^4} \int d\mathbf{k} \sum_{m=-1}^{1} \left[ \langle |a_{1m}(k)|^2 \rangle c_{sv}^2(11m;k) \tau_{1m}(k,\omega) / [1+i\omega\tau_{1m}(k,\omega)] \right] \qquad (8)$$

with

$$\langle |a_{1m}(k)|^2 \rangle = \frac{N}{4\pi}\left(1 + \frac{\rho_0}{4\pi}(-1)^m h(11m;k)\right) \qquad (9)$$

where h(11m;k) is the Fourier transform of the total pair correlation function in terms of spherical harmonics. The relaxation time $\tau_{10}$ longitudinal (m=0) is given by

$$\tau_{10}^{-1}(k) = 2D_{R0}\left[\left[\frac{k_BT}{2\varsigma(\omega)D_{R0}} + p'(k\sigma)^2\right]\left(1 - \frac{\rho_0}{4\pi}c(110;k)\right)\right] \qquad (10)$$

Here $p' = D_T/2D_{R0}\sigma^2$, $D_T$ and $D_{R0}$ being translational and rotational diffusivity ($D_{R0} = k_BT/\varsigma_s$, $\varsigma_s$ is the Stokes' friction) that measure importance of translational diffusion in the polarisation relaxation of dipolar liquid. We found that translational diffusion of solvent molecules can reduce the rotational friction on solute.

But for molecular species, situation becomes more complex. One can define the interactions in a molecular species by reference interaction site model (RISM) and formulate a mode coupling theory based on RISM.

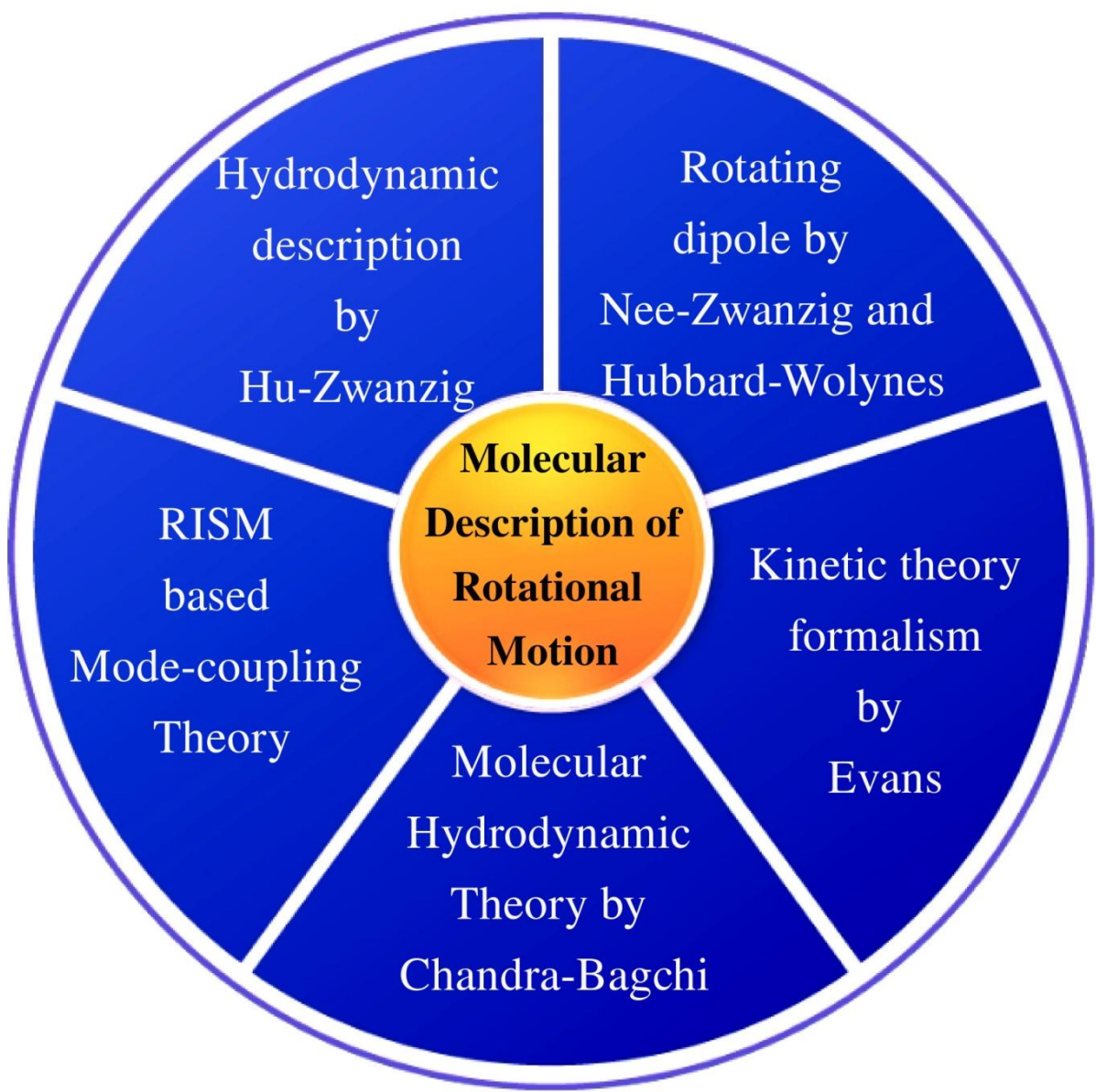

*Figure 1: Different theoretical formulations for the rotational friction*

In a series of papers, we recently addressed the translational diffusion of two polyatomic ions – nitrate and acetate – with emphasis on the former [33-35]. We demonstrated that in the case of polyatomic ions, the translational motion is coupled to its jump rotational motion which in turn was found to be coupled to the well-known jump dynamics of water molecules. Further, we demonstrated that symmetric ions like nitrate undergoes faster rotational jump motion in water which enhances its diffusivity whereas an asymmetric ion like acetate exhibit lower diffusivity with reduced rotational motion.

The orientational dynamics of polyatomic ions dominates the overall dynamics of the system. **Figure 2** shows the connections of different dynamical features with the orientational correlation function. Another interesting aspect of reorientational dynamics of polyatomic ions is rank ($\ell$) dependence of the orientational correlation function [36] [37]

$$C_\ell(t) = \frac{4\pi}{2\ell+1} \sum_{m=-\ell}^{\ell} \left\langle Y_{\ell m}^*(\Omega(0)) Y_{\ell m}(\Omega(t)) \right\rangle \tag{11}$$

where $Y_{\ell m}(\Omega)$ are the spherical harmonics of rank $\ell$. Generally, this correlation function can be defined in terms of their memory functions, in case of single particle it is $\zeta_R(z)$.

$$C_{\ell m}(z) = \left[ z + \frac{\ell(\ell+1)k_B T}{\zeta_R(z)} \right]^{-1} \tag{12}$$

where z is the Laplace frequency conjugate to time, t. Debye proposed the following simple expression for $C_{\ell m}(t)$ [38]

$$\begin{aligned} C_{\ell m}(t) &= C_{\ell m}(t=0) \exp\left[-\ell(\ell+1)D_R T\right] \\ \tau_\ell &= \left[\ell(\ell+1)D_R\right]^{-1} \end{aligned} \tag{13}$$

where rotational diffusion constant, $D_R$ is independent of rank $\ell$. But, $D_R$ can be dependent on rank $\ell$ due to non-Markovian effects and this is usually explained by the rank dependence of rotational dielectric friction coefficient, $\zeta_{R,DF}$ on $\ell$. In the Nee-Zwanzig [7] treatment they have neglected rank dependence, however, Hubbard-Wolynes [9] revised this theory including rank dependence and predicted inverse dependence of the dielectric friction on the value of $\ell$.

Now, different experiments probe dynamical response of the dipolar system at different ranks. Such as, Raman scattering and magnetic resonance experiment mainly probe the relaxation of orientational correlation of rank 2 ($C_2(t)$) whereas dielectric relaxation studies are connected with the collective orientational relaxation of rank 1 ($C_1(t)$). Debye's theory predicts that the ratio of $\tau_1/\tau_2$ should be 3. But depending on the polarisation interaction and rotational dynamics this ratio can vary.

In the polyatomic ionic system, where reorientational dynamics is predominated by jump rotational dynamics, the value of $\tau_1/\tau_2$ deviates significantly from 3. In this work, we have analysed rank dependence of orientational relaxation time and how it depends on the dielectric friction on the solute.

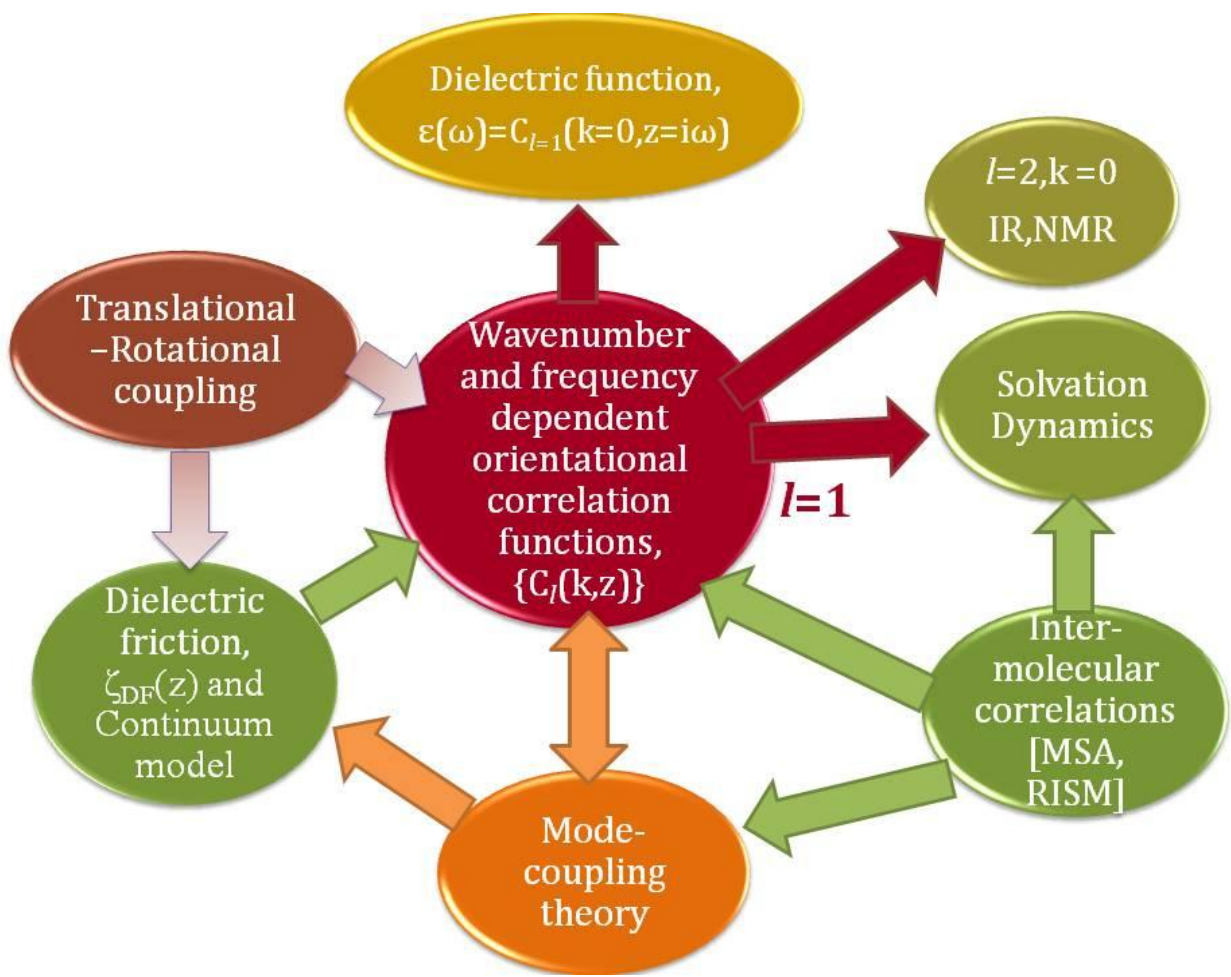

*Figure 2: Flowchart showing the connection of different dynamical properties with wavenumber and frequency dependent orientational correlation function.*

Since, Mode-Coupling Theory (MCT) predicts that amplitude of rotational friction depends on translational friction and vice versa, a self-consistent calculation would require detailed information about both rotational and translational friction. In fact, we shall see that there is symmetry into this inter-dependence within the MCT formalism, although the static inputs, like the direct correlation functions, enter at different levels. Reference interaction site model (RISM) provides detailed information about the solvent structure around the molecule. In addition, translational friction and rotational friction is found to be related with each other by a geometric factor that depends on the bond length and structure of the molecule.

Upto this point, we have mainly discussed about rotational dynamics. Now, most of the studies on translational diffusion of ions have been motivated by the well-known breakdown of Walden's product which states that the product of limiting ionic conductivity with solution viscosity should vary as inverse the size of the ion. The most dramatic departure from the Walden's rule is observed for rigid monatomic ions [39]. When plotted against inverse ion radius, the product of ionic conductivity and viscosity ($\lambda_0\eta_0$) shows a markedly non-monotonic size dependence.

There have been several schools of thought about this break-down. The initial approach using a solvent-berg picture was made redundant when it was found, by computer simulations, that the water molecules remain mobile even in the first hydration layer. An elegant approach was developed by Boyd [40] and Zwanzig [41], and Hubbard and Onsager (H-O) [42] who showed that a substantial contribution to the translational friction could arise from the solvent polarization forces that gave rise to additional energy dissipation . These authors used electro-hydrodynamic equations by including contributions from both electrostatics and hydrodynamics. However, H-O theory shows a good agreement with experiment up to intermediate ion size but failed to explain the behavior of small ions.

To overcome this drawback of continuum theories, a microscopic theory was first developed by Wolynes [2, 3, 43]. In this approach, translational dielectric friction on the ion was calculated from force-force time correlation function

$$\zeta = \frac{1}{3k_\text{B}T}\int_0^\infty dt \langle F(0) \cdot F(t) \rangle \tag{14}$$

where F(t) is the force exerted on the ion at time t due to ion-dipole interaction. $k_\text{B}$ is the Boltzmann constant and T is the absolute temperature. Later, this microscopic theory was further revised by our group by including ultrafast solvation dynamics into the dynamics of dipolar liquids using mode-coupling theory formalism [1, 44-47]. The resultant expression of the translational dielectric friction and rotational dielectric friction shows different size dependence.

In this article, we mainly focus to the rotational dielectric friction ($\zeta_{R,DF}$) of the polyatomic ions in water. At first, in section II, we discuss the theory of Alavi-Waldeck to measure $\zeta_{R,DF}$ numerically. We have simulated sulphate, nitrate and acetate ion in water in this paper. In section III, the simulation details have been discussed and then in section IV we have analysed first-rank and second rank orientational correlation function for all the ions and model systems we have used and present a comparison of the rotational dynamics of the system with debye relaxation. Then, we present a detailed atomistic simulation technique to calculate rotational dielectric friction on polyatomic ions in water. Finally, we compare the value of $\zeta_{R,DF}$ from Alavi-Waldeck theory and simulation. In the section IV.E. and IV.F.. we have developed a self-consistent mode-coupling theory approach to calculate rotational dielectric friction of such polyatomic systems in water using reference interaction site model (RISM) theory. Section IV.G. presents a detailed analysis of the microscopic solvation shell structure of these three ions and their uncharged analogue and aims to describe the poor agreement of Alavi-Waldeck treatment for sulphate ion with our simulation results.

## II. Dielectric Friction: The Generalized continuum model of Alavi and Waldeck(AW)

In a notable development, Alavi and Waldeck pointed out that most of the earlier theoretical studies suffered from an important draw-back which was the assumption of a point charge or point dipole embedded at the centre of a spherical or ellipsoidal cavity[8]. This precludes the application of the theories (continuum or molecular) to molecular ions and large molecules like proteins. To understand the importance of this observation, let us consider the case of sulphate ion. In this case, the negatively charged oxygen atoms of $SO_4^{2-}$ are close to the surface, and as a result, water molecules, themselves consisting of charged atoms, experience much stronger force from the oxygen atoms of the sulphate ion. The reverse is also true. In such a case, approximation of the sulphate ion (or proteins with charged groups on the surface) as a point dipole at the centre is bound to breakdown.

AW presented a generalization of Fatuzzo-Mason-Nee-Zwanzig treatment that, although followed the general method of NZ, is considerably more involved. For arbitrary multiple charge distribution model, AW calculated dielectric friction from the expression:

$$\varsigma_{DF} = \left(\frac{8}{R_c}\right)\left(\frac{\varepsilon_S - 1}{(2\varepsilon_1 + 1)^2}\right)\tau_D \sum_{j=1}^{N}\sum_{i=1}^{N}\sum_{l=1}^{\infty}\sum_{m=1}^{l}\left(\frac{2l+1}{l+1}\right)\frac{(l-m)!}{(l+m)!}m^3 q_i q_j \left(\frac{r_i}{R_c}\right)^l \left(\frac{r_j}{R_c}\right)^l P_l^m(\cos\theta_i) P_l^m(\cos\theta_j) \cos m\varphi_{ji}$$
(15)

where $P_l^m(x)$ are the Legendre polynomials, $R_c$ is the cavity radius, ($r_i$, $\theta_i$, $\varphi_i$) are the polar coordinates of the charges, $\varphi_{ji} = \varphi_j - \varphi_i$, $q_i$ is the partial charge of ith atom, $\varepsilon_S$ is the static dielectric constant of the solvent and $\tau_D$ is the Debye relaxation time.

They have considered a rigid stationary point charge, $q_i$ is at position $r_i$ inside a cavity (radius $R_c$) ($r_i \leq R_c$) on the +z axis. The electrostatic potential inside and outside the cavity is given by

$$\Phi_{out}(r,\theta) = \sum_{L=0}^{\infty} B_l \frac{P_l(\cos\theta)}{r^{l+1}}$$
(16)

$$\Phi_{in}(r,\theta) = \frac{q_i}{|r - r_i|} + \sum_{l=0}^{\infty} A_l r^l P_l(\cos\theta)$$
(17)

Inside the cavity, the first term in (17) comes from potential of the point charge and second term is the reaction potential due to the polarisation of the dielectric medium. Suitable boundary condition gives the reaction potential inside the cavity. For arbitrary coordinates, $r_i$, $\theta_i$, $\varphi_i$, when charge distribution is stationary, reaction potential for N point charges is written as

$$\Phi_{rxn}(r,\theta,\varphi) = \sum_{i=1}^{N}\sum_{l=0}^{\infty}\sum_{m=-l}^{l} \frac{-q_i r_i^l}{a^{2l+1}}\left[\frac{\varepsilon_s - 1}{\varepsilon_s + (l/(l+1))}\right] \times \frac{4\pi}{2l+1} Y_{lm}^*(\theta_i,\varphi_i) r^l Y_{lm}(\theta,\varphi)$$
(18)

Where $Y_{lm}$ are the spherical harmonics and $\varepsilon_s$ is the static dielectric constant of the medium. The negative gradient of the reaction potential gives the reaction field, **R**.

The torque exerted on a dipole by an electric field **E** is

$$|N| = |\mu \times E| = \mu E \sin\phi \tag{19}$$

Now, for the rotation of the charge distribution with frequency $\omega$, electric field oscillates and in the reaction field, $R(\omega)$, the torque on the rotating dipole is given by

$$|N(\omega)| = \mu |R(\omega)| \sin\phi(\omega) \tag{20}$$

where $\varphi(\omega)$ is the phase lag angle. Now, the frequency dependent friction ($\zeta(\omega)$) is obtained as the torque divided by the angular frequency for the rotation about z axis

$$\zeta_{lm}(\omega) = \left| \frac{[N(\omega)]_{lm}}{\omega_z} \right| \tag{21}$$

This leads to the final form of frequency dependent dielectric friction for arbitrary charge distribution,

$$\zeta(\omega) = \sum_{j=1}^{N}\sum_{i=1}^{N}\sum_{l=0}^{\infty}\sum_{m=-l}^{l} \frac{-q_i q_j}{a\omega} \left(\frac{r_i}{R_c}\right)^l \left(\frac{r_j}{R_c}\right)^l \left|\frac{\varepsilon(m\omega)-1}{\varepsilon(m\omega)+(l/(l+1))}\right| \frac{4\pi}{2l+1} im Y_{lm}^*(\theta_i,\varphi_i) \\ \times Y_{lm}\left[\theta_j, \varphi_j + \varphi_l(m\omega)\right] \tag{22}$$

In the limit of slow reorientation, zero frequency friction is written as Eq. (15)

We have calculated dielectric friction on the three polyatomic ions, sulphate, nitrate, acetate in water by using Eq. (15). The maximum value of $l$ in the Legendre polynomial is taken as 50. As our calculation involves water as the medium, we have used the value of $\varepsilon_S$ 78 and the value of Debye relaxation time, $\tau_D$ as 8.3 ps. This method is extremely sensitive to the cavity radius chosen for the molecule. Here we have taken the cavity radii from the radial distribution function (RDF) data of water around the ions such that if the RDF starts from the distance x, then the ratio of x and $R_c$ is 0.75 for all the ions. The resultant value of $R_c$ is therefore close to the maxima of the first peak of RDF.

The formula (Eq. (15)) gives the rotational friction for the rotation around z axis. In case of molecular species, to define its rotation, we need to take different rotational axis. Therefore, we have selected three basis vectors as rotational axis for three molecular ions as shown in the **Figure 3**.

For nitrate ion, we have taken the perpendicular vector to the plane ($C_3$ axis) as $\vec{z}$, one N-O bond vector ($C_2$ axis) as $\vec{y}$ and the cross product of these two as $\vec{x}$. Similarly, in case of acetate ion we have taken C1-C2 bond vector ($C_2$ axis) as $\vec{z}$, the perpendicular vector to the plane of the ion as $\vec{y}$ and cross product of these two as $\vec{x}$. But, sulphate ion is not a planar ion. Therefore, we have chosen one S-O bond vector ($C_3$ axis) as $\vec{z}$ and selected two other basis vectors by Gram-Schmidt orthogonalisation.

We have calculated rotational friction of the molecules for all three rotations along three basis vectors. The averaged values of rotational dielectric friction from AW formula are tabulated below in **Table I**.

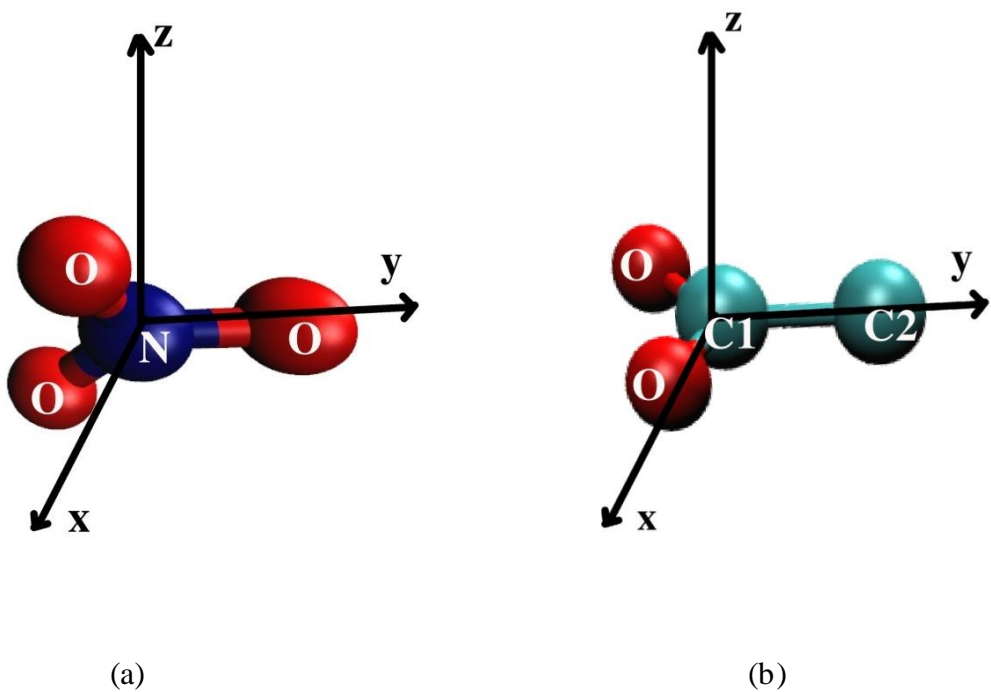

(a)  (b)

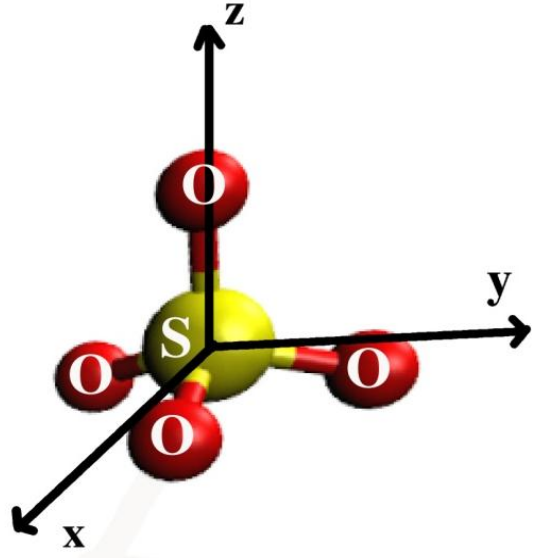

(c)

*Figure 3: Basis vectors for the calculation of angular velocity of all the three ions.*

*Table I: Dielectric friction for rotation of aqueous polyatomic ions calculated by AW formula*

| Ion | $\varsigma_{DF, AW}$ (in erg.s) |
|---|---|
| Nitrate | $1.63*10^{-24}$ |
| Acetate | $1.83*10^{-24}$ |
| Sulfate | $4.48*10^{-24}$ |

## III. Molecular Dynamics Simulation

Molecular dynamics simulations of potassium sulphate ($K_2SO_4$), potassium nitrate ($KNO_3$) and potassium acetate ($CH_3COOK$) in water have been carried out. For aqueous $KNO_3$ and aqueous $CH_3COOK$ we have used LAMMPS package[48] and for aqueous $K_2SO_4$ we have used the DL_POLY [49] package. Rigid non-polarizable force field parameters have been used for water as well as ions. SPC/E model[50] has been employed for water. For potassium ion, in all the simulations, the OPLS-AA[51] force field has been used. For nitrate ion, the potential model suggested by Vchirawongkwin *et al.* have been employed[52], for acetate ion, we have used united atom OPLS force field [51, 53] and for sulphate ion, the force filed developed by Cannon *et al.* has been used[54]. For aq. $KNO_3$ and potassium acetate, 16 cations and 16 anions and for aq. $K_2SO_4$, 16 cations and 8 anions have been taken in 8756 water molecules in a cubic simulation cell of length 64.04 Å. Simulations were carried out in the microcanonical ensemble with periodic boundary conditions with a cut-off radius of 15 Å. The long-range forces were computed with Ewald summation [55, 56]. Trajectory was propagated using a velocity Verlet integrator with a time step of 1 fs. The coordinates, velocities and forces were stored every 5 fs for subsequent use for the evaluation of various properties.

We have carried out a different set of simulations with all the anions, sulphate, nitrate and acetate in water but with zero partial charges on each constituent atom of the ions. System size, LJ interaction parameters for anions and simulation details are the same discussed above.

## IV. Results and Discussions

### A. Rotational relaxation

As here our main goal is to analyse rotational dynamics of the three different polyatomic ions in water (nitrate, acetate, sulphate), we first measure rotational relaxation time by computing first-rank and second-rank orientational time correlation function, $C_\ell(t)$ that is defined as:

$$C_\ell(t) = \left\langle P_\ell(\vec{n}(t_0) \cdot \vec{n}(t_0 + t)) \right\rangle \quad (23)$$

where $P_\ell(x)$ is the second order Legendre polynomial defined as

$$\begin{aligned} P_1(x) &= x \\ P_2(x) &= \frac{1}{2}(3x^2 - 1) \end{aligned} \quad (24)$$

Here $\vec{n}$ is the unit vector along the bond vectors of polyatomic ions.

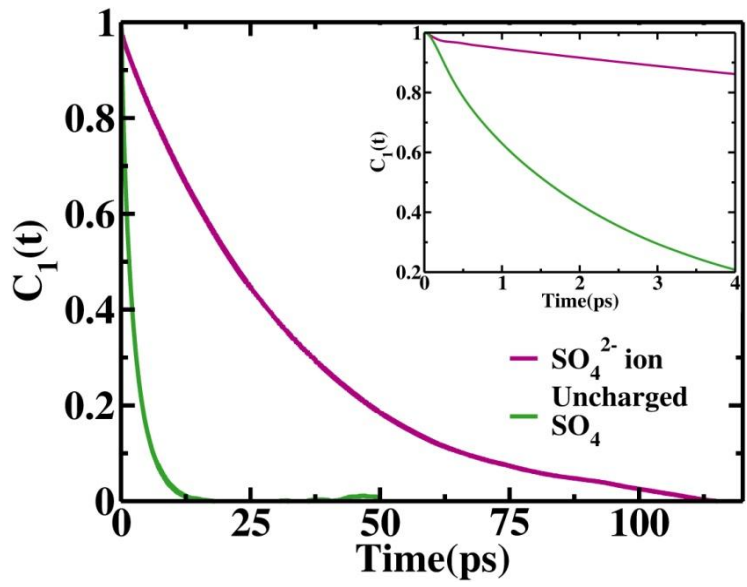

(a)

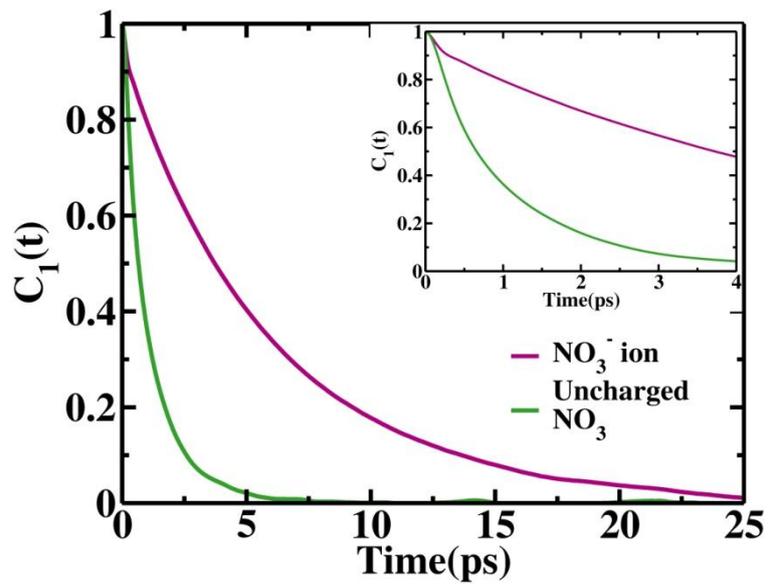

(b)

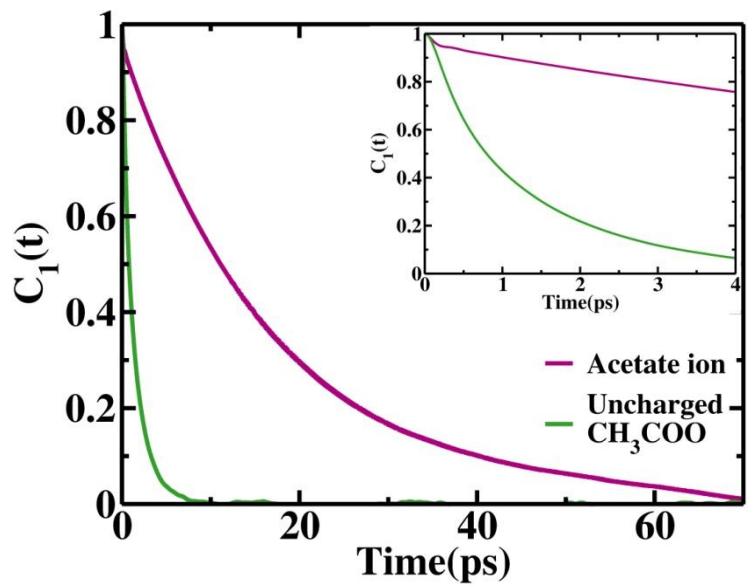

(c)

*Figure 4: First-order time correlation function, $C_1(t)$ of aqueous nitrate, acetate and sulphate ion and their uncharged analogues in water. The insets show the different nature of initial decay for charged and uncharged versions of the molecules.*

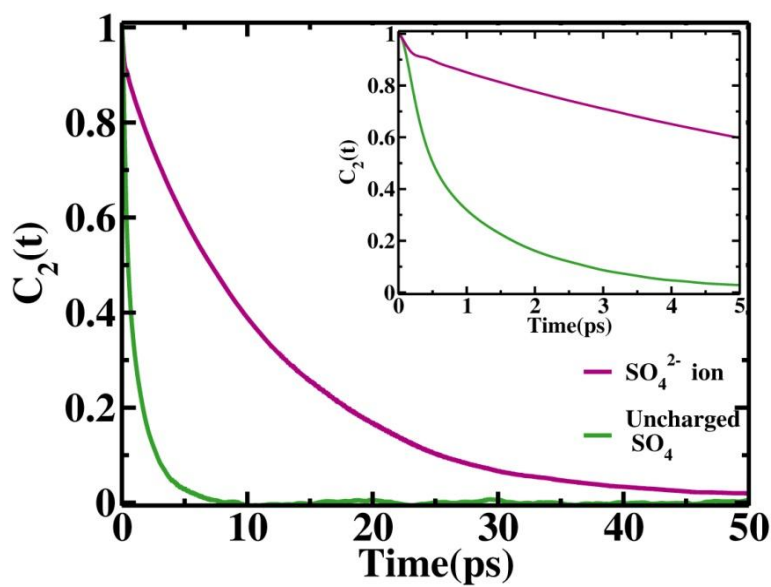

(a)

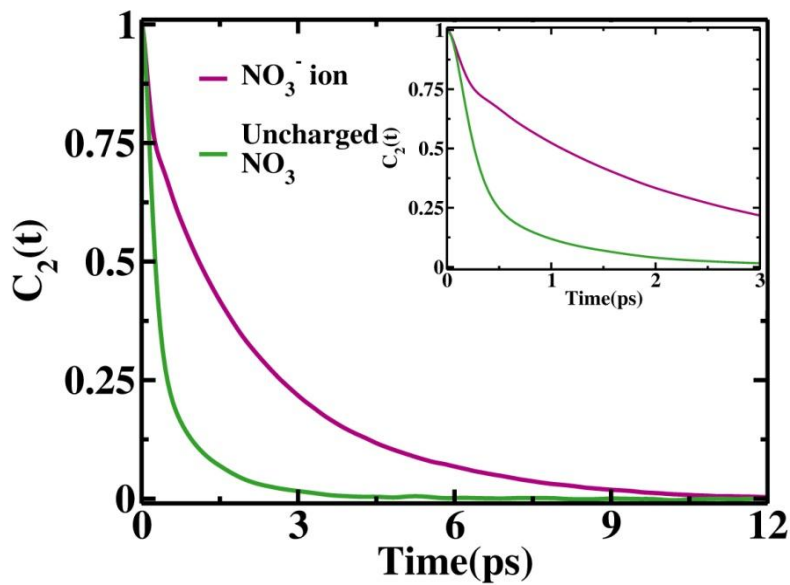

(b)

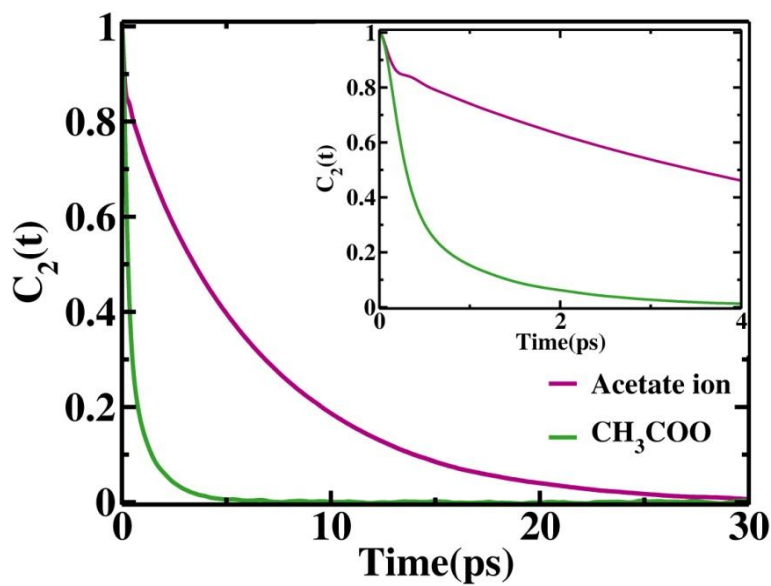

(c)

*Figure 5: Second-order orientational time correlation function, $C_2(t)$ of aqueous nitrate, acetate and sulphate ion and their uncharged analogues in water. The insets show the different nature of initial decay for charged and uncharged versions of the molecules.*

**Figure 4** and **Figure 5** show first-rank and second-rank rotational time correlation function of sulphate, nitrate and acetate ions respectively and also their uncharged analogues in water. For the charged ions, both $C_1(t)$ and $C_2(t)$ curves are best fitted by two exponential functions ($C_\ell(t) = a_1 \exp(-t/\tau_1) + a_2 \exp(-t/\tau_2)$), one of them ($\tau_1$) is responsible for librational motion and another ($\tau_2$) is for full rotation. The decay constants of $C_1(t)$ and $C_2(t)$ are listed in **Table II** and **Table III** respectively.

The relaxation of orientational time correlation function of water has been studied in detail in many previous works[57-59]. Similar to water, the rotation of these polyatomic ions (described by the decay of the relaxation of both $C_1(t)$ and $C_2(t)$ ) are dominated by diffusive motion, characterized by long diffusion time constant ($\tau_2$). These decay functions derive small contribution from the faster rotations of femtosecond timescale due to librational motion ($\tau_1$). First-rank correlation function relaxes much slower than the second-rank. Further, the relaxation of nitrate ion is the fastest and that of the sulphate ion is the slowest due to the higher partial charges and a highly structured solvation shell. On the other hand, as already pointed out in our previous paper, acetate ion also exhibit slower rotational dynamics than that of the nitrate ion due to its asymmetric charge distribution.

*Table II: Decay constants of first-rank orientational correlation function, $C_1(t)$ of the ions in water*

|  | $a_1$ | $\tau_1$(ps) | $a_2$ | $\tau_2$(ps) |
|---|---|---|---|---|
| **Nitrate ion** | 0.07 | 0.24 | 0.93 | 6.01 |
| **Acetate ion** | 0.05 | 0.165 | 0.95 | 17.3 |
| **Sulfate ion** |  |  | 1.0 | 29.7 |

*Table III: Decay constants of second-rank orientational correlation function, $C_2(t)$ of the ions in water*

|  | $a_1$ | $\tau_1$(ps) | $a_2$ | $\tau_2$(ps) |
|---|---|---|---|---|
| **Nitrate ion** | 0.211 | 0.214 | 0.789 | 2.35 |
| **Acetate ion** | 0.141 | 0.159 | 0.859 | 6.498 |
| **Sulfate ion** | 0.08 | 0.23 | 0.92 | 11.66 |

*Table IV: Comparison of ratio of first rank and second-rank orientational relaxation time*

*constants for nitrate, acetate and sulphate ions.*

|  | $\dfrac{\tau_{\ell=1}}{\tau_{\ell=2}}$ |
|---|---|
| Nitrate ion | 2.56 |
| Acetate ion | 2.66 |
| Sulfate ion | 2.55 |

**Table IV** shows the ratio of the two decay constants obtained from first-rank and second-rank correlation functions. The time constants are selected such that we get compare the diffusive behaviour of different ranks of time correlation function. It is evident from the values that rotational dynamics significantly deviates from Debye's behaviour that predicts the ratio $\tau_{\ell=1}/\tau_{\ell=2}$ to be 3. As mentioned earlier, the rotational dynamics of polyatomic ions involve jump reorientation. Many theoretical models have been proposed in the past to describe rotational jump diffusion, most notable among them are by Kubo[60, 61], Ivanov[62] and Anderson[63] and it was further revised by Seki *et. al.*[64].

The jump model of Ivanov[62] [57] suggests the rank dependence of relaxation time should follow the relation

$$\tau_\ell^{\text{jump}} = \tau_0 \left\{ 1 - \frac{1}{2\ell+1} \frac{\sin\left[(\ell+1/2)\Delta\theta\right]}{\sin(\Delta\theta/2)} \right\}^{-1} \tag{25}$$

where $\tau_0$ is the jump time and $\Delta\theta$ is the magnitude of the jump. In the previous work[33], we have computed the jump angle of nitrate ion as $110^0$ and $80^0$ in two different mechanisms of hydrogen bond switching. Using these values, Ivanov's jump model predicts a value of 1.7 for the ratio of $\tau_{\ell=1}^{\text{jump}}/\tau_{\ell=2}^{\text{jump}}$. But this Ivanov's ratio of rotational relaxation time involves only the jump reorientation and neglects the contribution of the slow diffusive reorientation. The latter should predict a value closer to three. The combination of these two contributions appears to be responsible for the larger value observed for the ratio from simulation.

Unlike the polyatomic ions discussed above, in case of uncharged molecules ($SO_4$, $NO_3$, $CH_3COO$), the initial decay is Gaussian in nature. This result perfectly matches with the rotational relaxation theory of Zwanzig for a planar Brownian rotator. Using Fokker-Planck equation, Zwanzig derived the expression for $l$-th order time correlation function for the Brownian rotator [65].

$$C_l(t) = \exp\left\{-\frac{kT}{I}\tau l^2\left[t-\tau(1-e^{-t/\tau})\right]\right\} \quad (26)$$

For short times (t<<τ), it has Gaussian nature ($C_l(t) \to \exp\left\{-\frac{kT}{I}\frac{1}{2}l^2t^2\right\}$) and for long times (t>>τ), it decays as an exponential function ($C_l(t) \to \exp\left\{-\frac{kT\tau}{I}l^2t\right\}$) and the crossover between two behaviour is at t=τ.

Therefore, in the case of uncharged molecules in water, we have fitted both the $C_1(t)$ and $C_2(t)$ curves with the fitting function $C_\ell(t) = a_1\exp(-(t/\tau_1)^2) + a_2\exp(-t/\tau_2)$ having one Gaussian and one exponential function. The decay constants corresponding to $C_1(t)$ and $C_2(t)$ are listed in **Table V** and **Table VI** respectively. These curves decay much faster than corresponding curves for charged ions which dictates the enhancement of rotational motion in absence of dielectric friction for rotation in these uncharged molecules in water.

*Table V: Decay constants of first-order orientational correlation time-correlation function, $C_1(t)$ of the uncharged analogues of nitrate, acetate and sulphate in water*

|  | a₁ | τ₁(ps) | a₂ | τ₂(ps) |
|---|---|---|---|---|
| Uncharged NO₃ | 0.23 | 0.46 | 0.77 | 1.29 |
| Uncharged CH₃COO | 0.22 | 0.496 | 0.78 | 1.59 |
| Uncharged SO₄ | 0.13 | 0.602 | 0.87 | 2.83 |

*Table VI: Decay constants of second-order orientational correlation time-correlation function,*

*$C_2(t)$ of the uncharged analogues of nitrate, acetate and sulphate in water*

|  | $a_1$ | $\tau_1$(ps) | $a_2$ | $\tau_2$(ps) |
|---|---|---|---|---|
| Uncharged $NO_3$ | 0.58 | 0.25 | 0.42 | 0.84 |
| Uncharged $CH_3COO$ | 0.55 | 0.27 | 0.45 | 0.99 |
| Uncharged $SO_4$ | 0.37 | 0.34 | 0.63 | 1.51 |

*Table VII: Comparison of ratio of first rank and second-rank orientational relaxation time constants for the uncharged molecules in water*

|  | $\dfrac{\tau_{\ell=1}}{\tau_{\ell=2}}$ |
|---|---|
| Uncharged $NO_3$ | 1.54 |
| Uncharged $CH_3COO$ | 1.61 |
| Uncharged $SO_4$ | 1.87 |

**Table VII** lists the ratio of the decay constants for full rotations corresponds to $C_1(t)$ and $C_2(t)$ for the uncharged analogues of the three ions in water. In these cases, rotational dielectric friction vanishes for the uncharged solutes. As a result, first-rank time correlation function decays much faster due to faster rotational motion of the uncharged molecules. This causes further deviation of the value of this ratio of time constants from 3. In this case we can write the relaxation equation as a sum over three contributions: Gaussian nature of relaxation due to inertial motion, diffusive behaviour as described by Debye and jump rotation described by Ivanov-Anderson(IA).

$$C_\ell(t) = a_1 \exp\left[-\ell^2(t/\tau_G)\right] + a_2 \exp\left[-\ell(\ell+1)(t/\tau_D)\right] + a_3 \exp\left[-f_{IA}(\ell)(t/\tau_{Jump})\right] \qquad (27)$$

It is evident from the time constants listed above that in case of uncharged molecules the contribution from inertial motion to the overall relaxation is significant compared to the case of charged ions in water. However, an explicit evolution of the rank dependence has not been possible yet as we cannot have a proper decomposition of the coefficients ($a_n$). Therefore, the dynamics involves more complex rank dependence.

## B. Rotational diffusivity of Aqueous Polyatomic ions in water

The rotational motion of the polyatomic ions can also be quantified in terms of the angular velocity autocorrelation function that is defined as

$$C_\omega(t) = \langle \omega(0) \cdot \omega(t) \rangle \quad (28)$$

where ω(t) is the angular velocity vector at time t. Now, using this autocorrelation function, the rotational diffusion coefficient can be calculated from Green-Kubo formula[66]

$$D_R = \frac{k_B T}{I} \int_0^\infty \frac{C_\omega(t)}{C_\omega(t=0)} dt \quad (29)$$

where I is the mass moment of inertia of the particle which is determined by adding the moment of each atom within the polyatomic ions about the centre of mass. Here, the main challenge is to calculate angular velocity of these molecular ions properly.

As our simulation involves all the ions and water molecules as classical rigid bodies, we can use the conventional quaternion formulation, $\mathbf{q}\{\chi, \eta, \xi, \varsigma\}$ to describe the orientation of these polyatomic ions. At first, we define three suitable orthonormal basis vectors for these three polyatomic ions (as shown in Figure 3) which are known as "body-fixed frame".

We obtain rotational matrix with the direction cosines of the body fixed axes and from the matrix elements we compute the quaternions. Finally, the time derivatives of quaternions, $\dot{q}(t)$ can be expressed in terms of angular velocity components along three basis vectors [67][68][69] ($\omega_x$, $\omega_y$, $\omega_z$)

$$\dot{q}(t) = \begin{pmatrix} \dot{\xi} \\ \dot{\eta} \\ \dot{\varsigma} \\ \dot{\chi} \end{pmatrix} = \frac{1}{2} \begin{pmatrix} -\varsigma & -\chi & \eta & \xi \\ \chi & -\varsigma & -\xi & \eta \\ \xi & \eta & \chi & \varsigma \\ -\eta & \xi & -\varsigma & \chi \end{pmatrix} \begin{pmatrix} \omega_x \\ \omega_y \\ \omega_z \\ 0 \end{pmatrix} \quad (30)$$

Here ξ, η, ζ, χ are the four quaternions. $\omega_x$, $\omega_y$ and $\omega_z$ are the principle angular velocity components in the body fixed frame. Finally we compute rotational diffusion with Eq. (28) and (29).

We have calculated rotational diffusion constant of all three polyatomic ions with charged and without charges and the values are listed below in **Table VIII**.

*Table VIII: Rotational diffusivities of charged ions and their uncharged analogues in water.*

| Ion | $D_{R,charged}(ps^{-1})$ | $D_{R,Uncharged}(ps^{-1})$ |
|---|---|---|
| Nitrate | 0.056±0.0024 | 0.07±0.003 |
| Acetate | 0.02±0.003 | 0.092±0.008 |
| Sulfate | 0.01±0.002 | 0.126±0.0045 |

## C. Simulation result of Dielectric friction of polyatomic ions in water

In the previous section, we have presented the results of rotational diffusion of charged ions and their uncharged analogues in water. **Table VIII** clearly shows that charged molecules has much lower rotational diffusivity than the uncharged molecules in water which arises for the absence of rotational dielectric friction in uncharged molecules that hinders the rotation of charged species. Now, we can calculate total friction acting on the species by their rotational diffusivity values using Stokes-Einstein relation. Therefore, we can measure the rotational dielectric friction for the polyatomic ions in water by subtracting the friction of unchanged species from the charged ones.

$$\varsigma_{DF} = \varsigma_{charged\_ion} - \varsigma_{Uncharged\_molecule} = k_B T \left[ \frac{1}{D_{charged\_ion}} - \frac{1}{D_{Uncharged\_molecule}} \right] \quad (31)$$

**Table IX** shows a comparison between the dielectric friction for rotation of all three polyatomic ions in water from the simulation and from the formula of Alavi-Waldeck. It shows a very good agreement between $\varsigma_{DF}$ and $\varsigma_{DF,AW}$ for nitrate and acetate. But, for sulphate, the agreement is not good.

Table IX: Comparison of rotational dielectric friction calculated from our simulation ($\varsigma_{DF}$) and Alavi-Waldeck's formula($\varsigma_{DF,AW}$)

| Ion | $\varsigma_{DF}$(erg.S) | $\varsigma_{DF,AW}$(erg.S) |
|---|---|---|
| Nitrate | $1.44*10^{-24}$ | $1.63*10^{-24}$ |
| Acetate | $1.57*10^{-24}$ | $1.83*10^{-24}$ |
| Sulfate | $3.69*10^{-24}$ | $4.48*10^{-24}$ |

Sulfate ion with higher charge than nitrate and acetate ions shows much difference between the cavity size when it has charge and when it does not have. But the decomposition of total friction into bare friction and dielectric friction ( Eq. (2)) assumes that the species will be in a similar environment. Later, we present the microscopic salvation structure analysis that reveals that the huge change in cavity radius from charged sulphate ion to uncharged $SO_4$ molecule results in the poor agreement of $\varsigma_{DF}$ with $\varsigma_{DF,AW}$.

## D. Translational Diffusion of charged and uncharged ions

We have computed translational diffusivity of all the species in water by calculating mean-square displacement of the molecule. The agreement with these diffusivity values of charged ions with the experimental diffusivity validates the forcefield parameters we have used for the simulation[70]. Translational diffusivity of uncharged molecules shows much higher diffusivity (~40-45%) than the corresponding charged ions for the similar reason of the absence of dielectric friction for translation as discussed in the previous section for rotational diffusion.

We have also calculated the corresponding diffusivity values from Stokes-Einstein (SE) relation using slip ( $D = \frac{k_B T}{4\pi\eta R}$ ) and stick boundary condition ( $D = \frac{k_B T}{6\pi\eta R}$ ). Slip boundary condition resembles to the uncharged species and stick boundary condition to the charged polyatomic ions. Here the ratio of $D_{charged}$ and $D_{uncharged}$ (obtained from simulation) is very close to the diffusivity obtained from stick and slip boundary condition for all the cases. But, the deviation is highest in nitrate ion as it exhibits fastest rotational dynamics and higher magnitude of translation-rotational coupling.

*Table X: Translational diffusivities of charges ions and uncharged molecules in water, the diffusivity predicted by Stokes-Einstein's relation using slip and stick boundary condition and the comparison with calculated diffusivity values*

| Ion | $D_{Charged}$ (*$10^{-5}$ cm$^2$/S) | $D_{Uncharged}$ (*$10^{-5}$ cm$^2$/S) | $D_{Stick}$ (*$10^{-5}$ cm$^2$/S)) | $D_{Slip}$ (*$10^{-5}$ cm$^2$/S) | $\frac{D_{Charged}}{D_{Uncharged}}$ | $\frac{D_{Stick}}{D_{Slip}}$ |
|---|---|---|---|---|---|---|
| **Nitrate** | 1.67 | 2.13 | 1.234 | 1.851 | 0.784 | 0.666 |
| **Acetate** | 1.094 | 1.74 | 1.27 | 1.908 | 0.629 | 0.666 |
| **Sulfate** | 0.95 | 1.34 | 1.13 | 1.698 | 0.708 | 0.666 |

### E. Mode Coupling Theory Approach

Mode-coupling theory appears to be a promising approach to deal with rotational friction ($\zeta_{R,DF}$) of polyatomic ions in water. As we have already discussed, using Kirkwood's formula for rotational friction, we can obtain $\zeta_{R,DF}$ from torque-torque time-correlation function(TTTCF) (Eq. (7)). Now, for a molecule like water, we can denote the orientation of the molecule by a point dipole and through the calculation of torque on the dipole-moment vector one obtains the expression for the memory functions of $\varsigma_{R,DF}$ using (Eq. (7))[12][71].

We can assume that the torque on a molecule arises due to density fluctuations. The orientation density of the point dipole can be expanded in spherical harmonics as

$$\rho(\boldsymbol{k},\boldsymbol{\Omega},t) = \sum_{l,m} a_{lm}(\boldsymbol{k},t) Y_{lm}(\boldsymbol{\Omega}) \tag{32}$$

Now, there are two types of orientational correlation function: the collective and the single-particle. The collective orientational correlation function can be written in k space where the collective limit signifies k=0 limit

$$C_{lm}(\boldsymbol{k},t) = \langle a_{lm}(-\boldsymbol{k},t=0) a_{lm}(\boldsymbol{k},t) \rangle \tag{33}$$

On the other hand, the single-particle orientational correlation function is written in terms of spherical harmonics

$$C_{lm}^{s}(t) = \langle Y_{lm}(\Omega_i(0)) Y_{lm}(\Omega_i(t)) \rangle \tag{34}$$

Classical density functional theory provides an elegant expression of torque on the point dipole due to density fluctuation

$$\boldsymbol{N}(\boldsymbol{r},\boldsymbol{\Omega},t) = k_B T \nabla_\Omega \int d\boldsymbol{r}' \, c(\boldsymbol{r}-\boldsymbol{r}',\boldsymbol{\Omega}) \, \delta\rho(\boldsymbol{r}',t) \tag{35}$$

where $\delta\rho(\boldsymbol{r},t) = \rho(\boldsymbol{r},t) - \rho_0$ is the density fluctuation around the average density $\rho_0$. $c(\boldsymbol{r}-\boldsymbol{r}',\Omega)$ is the direct correlation function between the point dipole at $(r,\Omega)$ and the solvent density at $r'$. The direct correlation function can be expanded in spherical harmonics to obtain

$$\boldsymbol{N}(\boldsymbol{r},\Omega,t) = \frac{1}{(2\pi)^3} \langle \nabla_\Omega Y_{lm}(\Omega) \rangle \int d\boldsymbol{k} \, e^{i\boldsymbol{k}\cdot\boldsymbol{r}} c_{lm}(\boldsymbol{k}) \delta\rho(\boldsymbol{k},t) \tag{36}$$

Now, the rotational friction for density fluctuation (dielectric friction) can be obtained as TTTCF using Kirkwood's formula

$$\varsigma_{R,DF} = \frac{1}{2k_B T} \int_0^\infty dt \, \frac{1}{4\pi V} \int d\boldsymbol{r} \, d\Omega \langle \boldsymbol{N}(\boldsymbol{r},\Omega,0) \cdot \boldsymbol{N}(\boldsymbol{r},\Omega,t) \rangle \tag{37}$$

The final expression for the memory functions of the rotational friction for the single-particle and the collective reorientational dynamics are written as

$$\Gamma_s(z) = \Gamma_{bare} + A\int_0^\infty dt\, e^{-zt} \int_0^\infty dk\, k^2 \sum_{l_1 l_2 m} c^2_{l_1 l_2 m}(k) F_{l_2 m}(k,t) \tag{38}$$

$$\Gamma_c(z) = \Gamma_{bare} + A\int_0^\infty dt\, e^{-zt} \int_0^\infty dk\, k^2 \sum_{l_1 l_2 m} F^s_{l_1 m}(k,t) c^2_{l_1 l_2 m}(k) F_{l_2 m}(k,t) \tag{39}$$

where the constant $A = \dfrac{\rho}{2(2\pi)^2}$. $c_{l_1 l_2 m}(k)$ is the $l_1 l_2 m$ th coefficient of the direct correlation function between solute and solvent and $F^s_{l_1 m}(k,t)$ and $F_{l_2 m}(k,t)$ are self and cross terms of orientational correlation function, defined as

$$F^s_{l_1 m}(k,t) = \left\langle e^{ik\cdot(r_i(t)-r_i(0))} Y_{lm}(\Omega_i(0)) Y_{lm}(\Omega_i(t)) \right\rangle \tag{40}$$

$$F_{l_2 m}(k,t) = \sum \left\langle e^{ik\cdot(r_j(t)-r_i(0))} Y_{lm}(\Omega_i(0)) Y_{lm}(\Omega_j(t)) \right\rangle \tag{41}$$

But for polyatomic ions, specially for those with zero dipole moment like nitrate, sulphate, situation becomes more complex. We can take the dipole moment vector through each bonds of such molecule using the microscopic expression for the torque on $i^{th}$ bond dipole moment vector at a position ($r_i$) and orientation ($\Omega_i$) at time t for a infinitesimal small rotation by an angle $\Theta_i$.

$$N_i(t) = -\nabla_{\Theta i} V_{eff,i}(r_i, \Omega_i, t) \tag{42}$$

with the effective potential

$$V_{eff,i}(r_i, \Omega_i, t) = -k_B T \int dr' d\Omega' c(r_i - r', \Omega_i, \Omega') \delta\rho(r', \Omega', t) \tag{43}$$

Let us assume, the unit vector for $i^{th}$ bond dipole moment at time t is $u_i(r,t)$. The rotational angle is then obtained as

$$\Theta_i(\delta t) = \cos^{-1}\left(u_i(r_1,t) \cdot u_i(r_2, t+\delta t)\right) \tag{44}$$

Now, for this change in orientation by $\Theta_i$, the torque on this bond is obtained as

$$N_i(t) = -\frac{\partial}{\partial \Theta_i} V_{eff,i}(r_i, \Omega_i, t) = -\frac{\partial r_i}{\partial \Theta_i}\frac{\partial}{\partial r_i} V_{eff,i}(r_i, \Omega_i, t) = -\frac{\partial r_i}{\partial \Theta_i}\frac{\partial}{\partial r_i}\left[\int dr' d\Omega' c(r_i - r', \Omega_i, \Omega') \delta\rho(r', \Omega', t)\right]$$

$$= k_B T A_{G,i} \frac{\partial}{\partial r_i}\left[\int dr' d\Omega' c(r_i - r', \Omega_i, \Omega') \delta\rho(r', \Omega', t)\right]$$

(45)

where $A_{G,i}$ is a geometric factor that varies from bond to bond and molecule to molecule. We discuss later the simple case of a rod-like solute molecule in liquid water where Eq. (45) can be simplified considerably.

Finally, the expression for effective torque density on the polyatomic system is then given by

$$\mathbf{N}(t) = \sum_{i=1}^{N_b} \mathbf{N}_i(t) \tag{46}$$

where $N_b$ is the total number of bond vectors in the polyatomic ion. Similarly, one can write the total force on a polyatomic ion or molecule (with $N_a$ number of atoms) as

$$\mathbf{F}(t) = \sum_{i=1}^{N_a} \mathbf{F}_i(t) \tag{47}$$

An interesting point here is the correlation between torques and forces at different sites can be correlated through correlations in the solvent. Thus, the total force-force time correlation function can be written for all combinations of sites

$$\langle \mathbf{F}(0) \cdot \mathbf{F}(t) \rangle = \sum_{(i,j)} \langle \mathbf{F}_i(0) \cdot \mathbf{F}_j(t) \rangle \tag{48}$$

The same is applicable for torque-torque correlation function, $\langle \mathbf{N}(0) \cdot \mathbf{N}(t) \rangle$. The above expression can be further decomposed into pure and cross-terms, the later contain force correlations at different sites.

$$\langle \mathbf{F}(0) \cdot \mathbf{F}(t) \rangle = \sum_{i} \langle \mathbf{F}_i(0) \cdot \mathbf{F}_i(t) \rangle + \sum_{(i \neq j)} \langle \mathbf{F}_i(0) \cdot \mathbf{F}_j(t) \rangle \tag{49}$$

The cross-correlations are expected to be smaller than the pure terms. If we neglect the crossterms, we have the pure term on one atom multiplied by the number of atoms. For example, for nitrate ion, we shall have four such terms. Neglect of the cross terms in the force on different atoms and keeping only the pure term is known as Kirkwood-Riseman approximation[72]. The entire discussion made above is applicable to torque terms as well.

The above formulation can be used to obtain both the fluctuating force and the fluctuating torque on a molecule. However, one needs the information about the site-specific direct correlation function and also the density fluctuation spectrum of the dipolar liquid. For polyatomic ions in liquid water, this implies detailed information about correlations.

One can use reference interaction site model (RISM) [73] to define the interaction on such molecular species and formulate a mode coupling theory based on RISM. Using RISM, we can express site-site pair correlation functions as

$$h_{\nu\nu'}^{ij}(r) = w_{\nu\beta}^{ii}(r) * c_{\beta\xi}^{ij}(r) * (w_{\xi\nu'}^{jj}(r) + \rho_\upsilon h_{\xi\nu'}^{jj}(r)) \tag{50}$$

where i and j denote two molecules and $\nu$ and $\nu'$ represent the sites on them. $h_{\nu\nu'}^{ij}(r)$ is the total correlation function of site $\nu'$ of j molecule around site $\nu$ of i molecule, $c_{\beta\xi}^{ij}(r)$ is the site-site direct correlation between site $\beta$ of i molecule and site $\xi$ of j molecule, $w_{\xi\nu'}^{jj}(r)$ is the intramolecular matrix of molecule j that depends on the site-separation of $\xi$ and $\nu'$, $l_{\xi\nu'}^{\upsilon}$. $\rho_\upsilon$ is the number density of solvent molecules and * signifies convolution. Here we assume molecule i is at position r and molecule j is at r' in the space fixed frame and we sum over all the contributions from different pairs of sites $\nu$ of the ith molecule and $\nu'$ of the jth molecule in the body fixed frame.

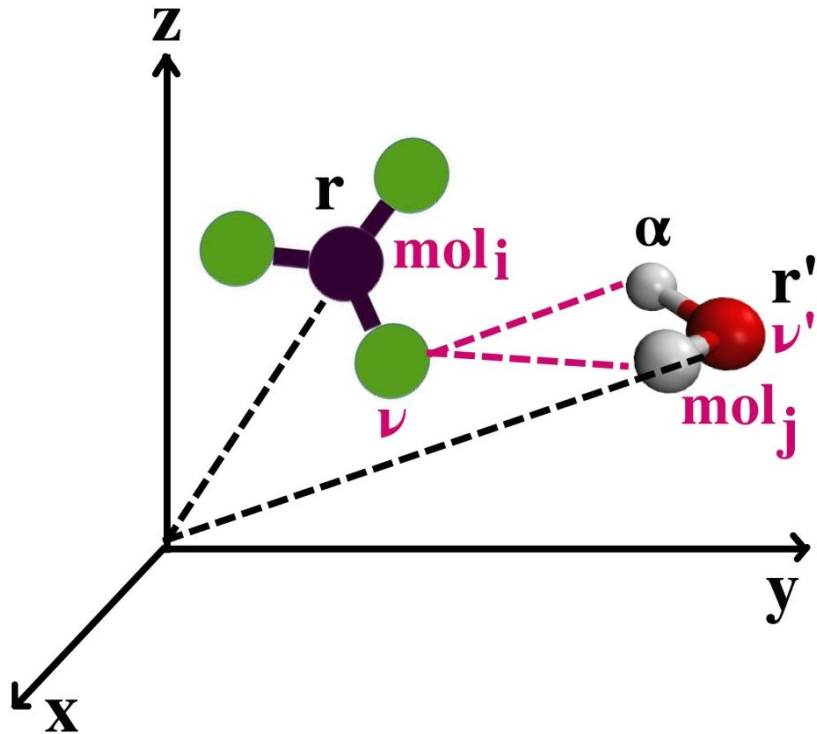

*Figure 6: Reference interaction site model (RISM) representation for nitrate ion and water*

Now, in 1D-RISM, the final expression is averaged over both solute and solvent orientational degrees of freedom and we get distribution function of solute-solvent pairs with only one variable, distance between their centre-of-mass. But, for an asymmetric molecule this does not give the full description of the solvation shell. However, in 3D-RISM, we obtain the distribution function of solvent interaction sites, $v'$ around solute molecule in three dimensional space position, **r** [74] [75]

$$h_{v'}^{ij}(\mathbf{r}) = \sum_{\alpha} \int d\mathbf{r}' c_{\alpha}^{ij}(\mathbf{r}-\mathbf{r}') \left[ \omega_{\alpha v'}^{ij}(\mathbf{r}') + \rho_{\alpha} h_{\alpha v'}^{ij}(\mathbf{r}') \right] \tag{51}$$

Here, we need a closure equation that relates h and c to uniquely define direct correlation function between solute and solvent. Among all the closure relations, Kovalenko-Hirata (KH) closure is the most popular one for 3D-RISM. KH is a combination of Hypernetted Chain (HNC) and mean spherical approximation (MSA). HNC is applied for the solvent density depletion including in the repulsive core (g<1) and the latter is applied for the solvent density enriched region, i.e., maxima of RDF peaks.

$$g^{KH}(r_{12},\Omega_1,\Omega_2) = \begin{cases} \exp\left[-\beta u(r_{12},\Omega_1,\Omega_2) + h(r_{12},\Omega_1,\Omega_2) - c(r_{12},\Omega_1,\Omega_2)\right] & \text{for } g(r_{12},\Omega_1,\Omega_2) \leq 1 \\ 1 - \beta u(r_{12},\Omega_1,\Omega_2) + h(r_{12},\Omega_1,\Omega_2) - c(r_{12},\Omega_1,\Omega_2) & \text{for } g(r_{12},\Omega_1,\Omega_2) > 1 \end{cases} \tag{52}$$

Further, the intermediate scattering function, F(k,t) can be expressed as

$$F_{vv'}^{ij}(k,t) = \frac{1}{N} \left\langle \sum_i e^{-i\mathbf{k}\cdot\mathbf{r}_i^v(0)} \sum_j e^{-i\mathbf{k}\cdot\mathbf{r}_j^{v'}(t)} \right\rangle \tag{53}$$

where i and j refer to 2 molecules and v and v' refer to the interaction sites on molecules.

To illustrate the above points, let us now consider a rod-like molecule (like $CO_2$). In this case we have $\frac{\partial r}{\partial \Omega} = L$ where L is the bond-length of the molecule. In such a situation force-force correlation function and torque-torque correlation function we can write

$$\langle N(0) \cdot N(t) \rangle = A_G(L) \langle F(0) \cdot F(t) \rangle \tag{54}$$

where the value of $A_G(L)$ is decided by the bond length and geometry of the molecule. Here, our aim is to obtain the force on each sites of the molecule that acts as a torque and helps in the rotation of the molecule. In $CO_2$ the force acting on oxygen atoms can be separated in three components: one along the C-O bond vector and other two is perpendicular to it. The perpendicular forces basically generate torque to rotate the molecule and the other one component of the force is transferred to the central atom through the connecting bond and helps the molecule to translate. This leads to the translation-rotational coupling.

## F. Self Consistent MCT

In the memory function equation of rotational friction (Eq. (39)), the second term, dielectric friction term, $\zeta_{R,DF}$ is calculated from torque-torque correlation function and the solution needs a self-consistent treatment ( as it is defined is such a way, both RHS and LHS has the term $\zeta_{R,DF}$)

$$\zeta_{R,DF}(z) = \frac{2k_B T \rho_0}{3(2\pi)^2} A_G^2 \int_0^\infty dt\, e^{-zt} \int_0^\infty dk\, k^2 S_{ion}(k,t) \left| c_{id}^{eff}(k) \right|^2 S_{Solv}^{dd}(k,t) \tag{55}$$

where $c_{id}^{eff}(k)$ is the effective dielectric constant between polyatomic ion and dipolar medium that we obtain from RISM theory. $S_{ion}(k,t)$ is the dynamic structure factor of the ion and $S_{Solv}^{dd}(k,t)$ is the dynamic structure factor for the dipolar solvent. $A_G$ is the geometric factor as discussed in previous section. The self consistent scheme has mainly three steps as shown in **Figure 7**:

1) The dynamic structure factor of solute (polyatomic ion), $S^{ion}(k,t)$ is first calculated from the translational diffusivity, $D_T^{ion}$ of the solute, assuming Gaussian approximation.

2) $D_T^{ion}$ is obtained from the time dependent velocity autocorrelation function, $C_v(\tau)$.

3) The frequency dependent velocity autocorrelation function, $C_v(z)$ is now related to frequency dependent total friction, $\zeta_{R,DF}(z)$ through the generalized Einstein's relation and we can always obtain $C_v(\tau)$ from $C_v(z)$ numerically by inverse Laplace transform.

4) The final $S^{ion}(k,t)$ is used to calculate $\zeta_{R,DF}(z)$ by Eq. (55) and the iteration continues until it converges.

Therefore, $\zeta_{R,DF}$ of the polyatomic ion is calculated starting from the mean-square displacement(MSD) of the solute self-consistently. No other previous study have used self-consistent scheme to calculate friction by RISM based mode-coupling theory.

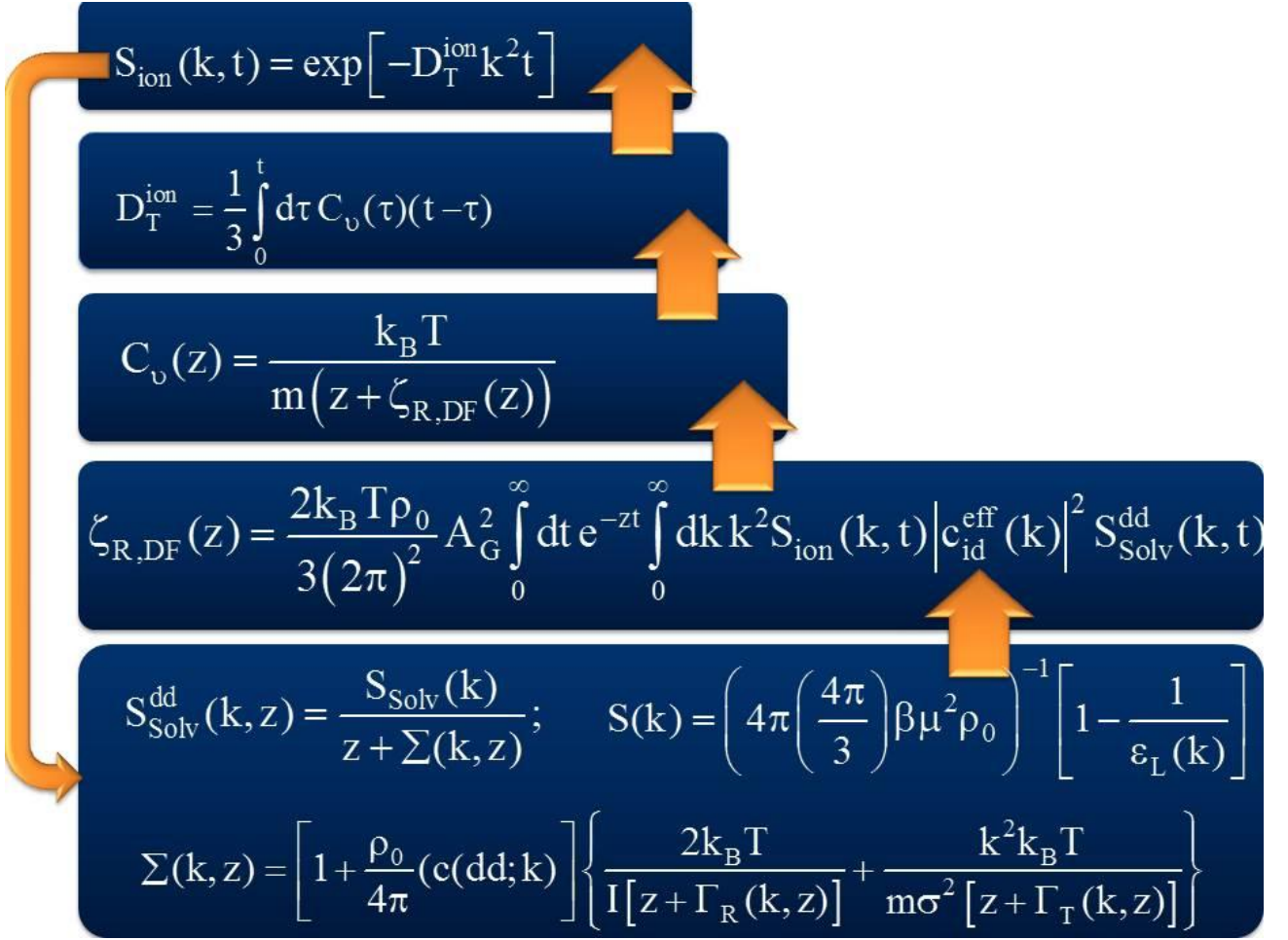

*Figure 7: Self-consistent scheme of Mode-coupling theory for the rotational dielectric friction of a polyatomic ion in water*

## G.   Analysis in Terms of Microscopic Structural Arrangements

In the mode-coupling theory, we have already discussed that the rotational dielectric friction, $\zeta_{R,DF}$ depends on the direct correlation function, c(r) (Eq. 55). Therefore, in this context, we analyze the structural rearrangement of water molecules around sulphate, nitrate and acetate ions from the simulation trajectory.

At first, we consider the simulation for the real ions (with charge) in water. **Figure 8**(a) and **Figure 8**(c) show the radial distribution function between the central atoms (carbon (C1) atom for acetate, nitrogen(N) for nitrate and sulphur(S) for sulfate) and the oxygen atoms (OW) of water molecules surrounding them. Although nitrate ion is symmetric unlike acetate ion, RDF of C1-OW and N-OW is more or less comparable (specially first peak). However, the number density, i.e. the peak height is not same as water molecules are not distributed symmetrically in acetate. In the region of two oxygen atoms of acetate ion, population of water is more but around carbon atom (C2) (shown in **Figure 3**) is very less.

But, compared to acetate and nitrate, sulfate ion shows quite different distribution of water surrounding it. As sulfate ion is much bigger (2.18 Å) than the nitrate (2.0 Å) and acetate ion (1.94 Å), peaks of $g_{S-OW}$ is shifted to higher r value. Also, as the sulfate ion is divalent, higher charge density can attract more water to form a compact hydration structure compared to the monovalent nitrate and acetate ions. However, the first peaks for O-OW pairs for all the ions coincides each other (**Figure 8(c)**).

We next calculate the partial structure factors, $S_{\alpha\beta}(k)$ for all these pairs from the corresponding radial distribution function, $g_{\alpha\beta}(r)$ (shown in **Figure 8(b)** and **Figure 8(d)**). The partial structure factor is defined as

$$S_{\alpha\beta}(k) = \delta_{\alpha\beta} + 4\pi\rho\left(c_\alpha c_\beta\right)^{1/2} \int_0^\infty \left[g_{\alpha\beta}(r) - 1\right] \frac{\sin kr}{kr} r^2 dr \qquad (56)$$

where α and β are two constituent atoms of solute and solvent respectively and $\alpha \neq \beta$. $c_\alpha$ and $c_\beta$ are defined as

$$\begin{aligned} c_\alpha &= \frac{N_\alpha}{N} \\ c_\beta &= \frac{N_\beta}{N} \end{aligned} \qquad (57)$$

Where $N_\alpha$ and $N_\beta$ are the number of α and β type of atoms in the system. **Figure 8(b)** and **Figure 8(d)** show the difference between structure factors for all these ions. Now, the structure factor between a solute solvent pair directly is related to the direct correlation function by Ornstein-Zernike equation.

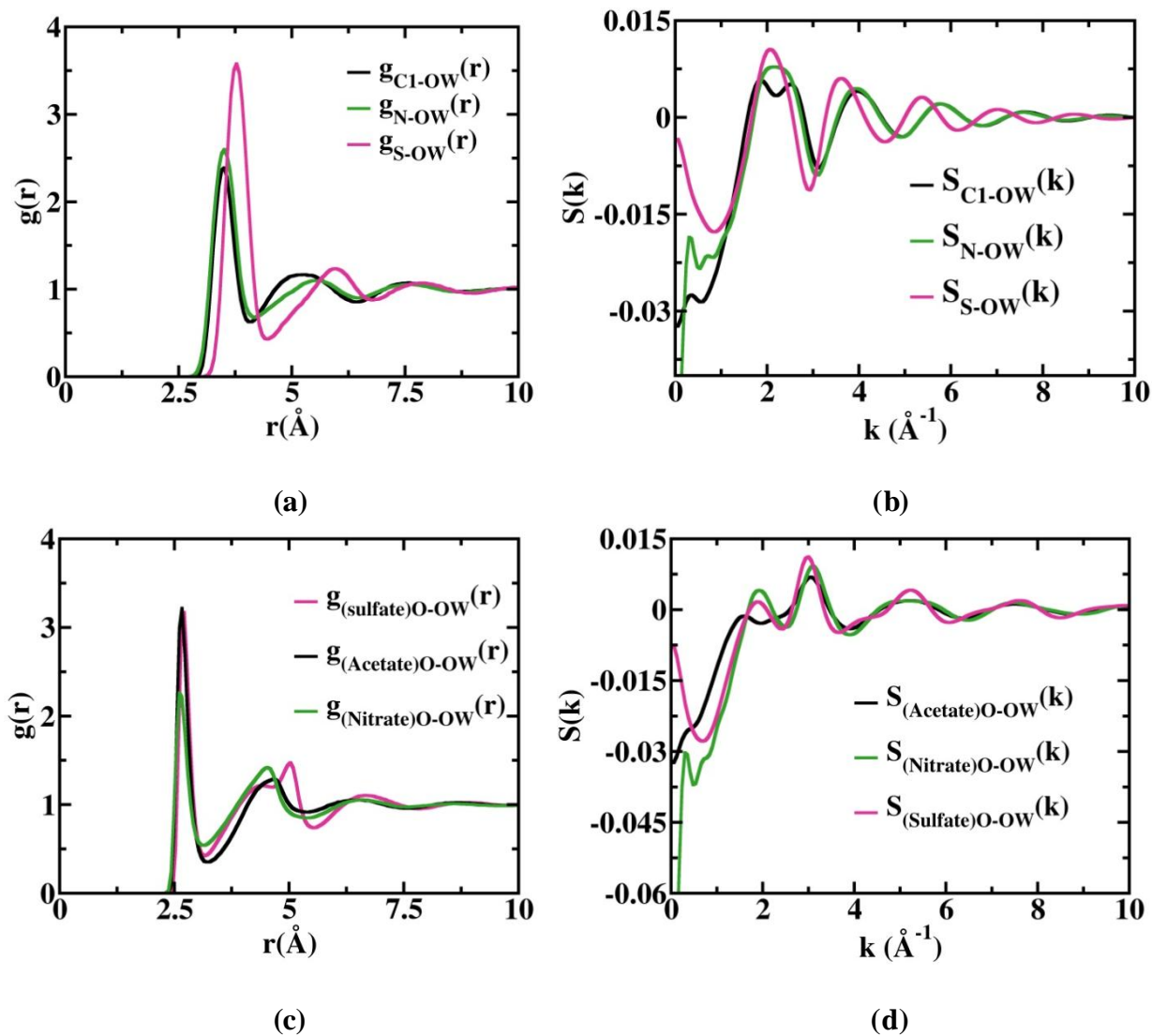

*Figure 8: Radial Distribution function between different pairs of atoms belongs to solute and solvent and partial structure factor for those pairs. We have denoted central carbon atom of acetate: C1, nitrate nitrogen: N, sulfate sulphur atom: S, oxygen atoms of acetate, nitrate and sulfate: O, water oxygen atoms: OW.*

Next, we want to analyse the change of the solvation shell structure in the other set of simulations with uncharged molecules. In the first set, we have simulated the real ions with accurate partial charges, whereas in the second set we have simulated $SO_4$, $NO_3$ and $CH_3COO$ molecule without any partial charges on the constituent atoms. *Figure 9* shows all the RDFs between different pairs of atoms of nitrate ion and water molecules. It clearly shows that solvation shell structure around nitrate ion breaks down in the absence of partial charges, peaks become broader and overlap each other.

*Figure 10* shows the RDFs between the different pairs of atoms of sulfate ions and water. Here the breakdown is much more pronounced as the ion has higher charge than the nitrate ion. Therefore, the structure of solvation shell in presence of partial charges on the ion and in case of uncharged ions are way more different. Comparing the position of RDF peaks from *Figure 10*(**a**) with *Figure 10*(**b**) and *Figure 10*(**c**) with *Figure 10*(**d**), it is evident that not only the structure breaks down here but the cavity size changes much more in this case when the charges are removed compared to the change of nitrate ion after removing charges. This result can explain the poor agreement of rotational dielectric friction of sulphate ion calculated from simulation with the Alavi-Waldeck theory (shown in *Table IX*).

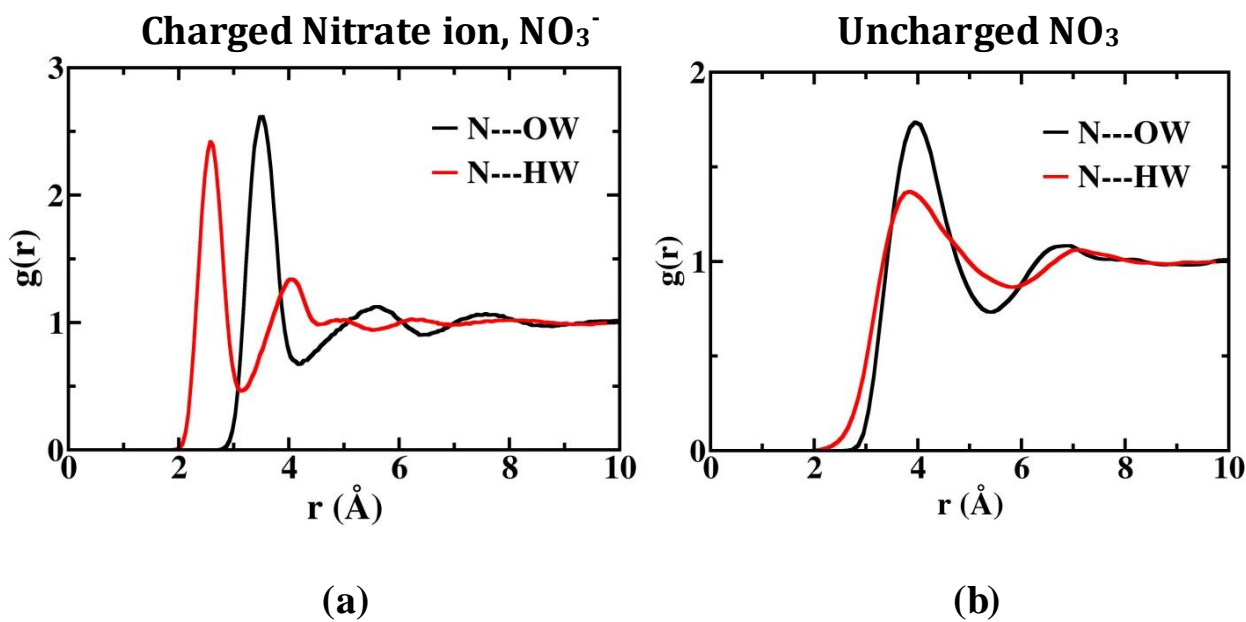

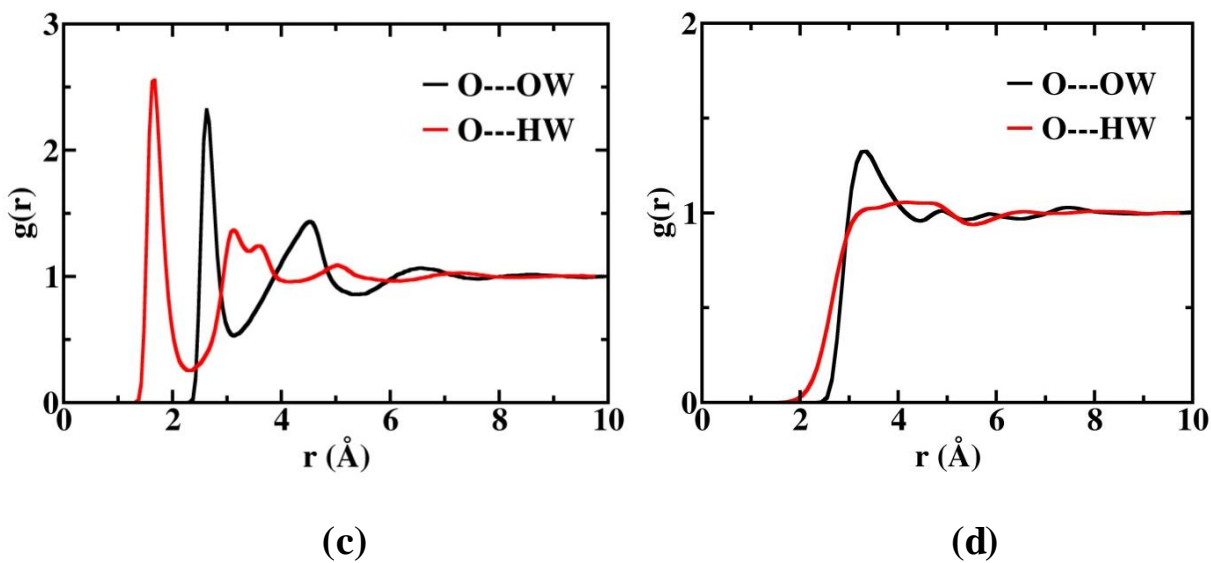

*Figure 9: Radial distribution function for water around (a,c) nitrate ions and (b,d) for uncharged $NO_3$ molecule in water. We have denoted nitrate nitrogen:N, nitrate oxygen:O, water oxygen: OW and water hydrogen:HW.*

**Charged sulfate ion, $SO_4^{2-}$**     **Uncharged $SO_4$**

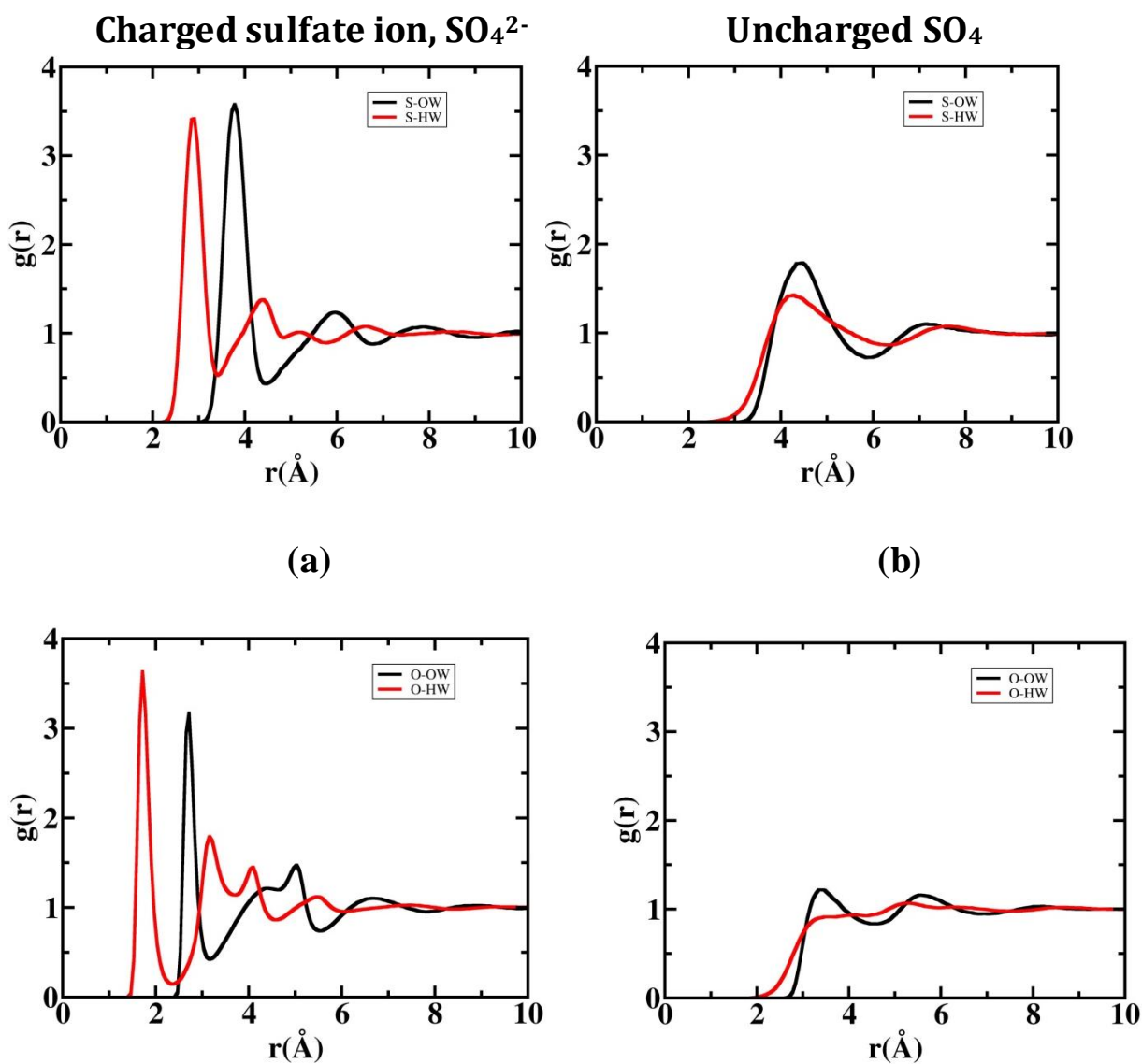

(a)     (b)

**(c)** **(d)**

*Figure 10: Radial distribution function for water around (a,c) sulphate ions and (b,d) for uncharged SO₄ molecule in water. We have denoted sulfate sulphur atom:S, sulfate oxygen atom:O, water oxygen: OW and water hydrogen:HW.*

## V. Conclusions

In polyatomic ions, the problem of translational diffusion (centre-of-mass movement) and rotational diffusion are entangled and cannot be separated from each other. In case of rigid alkali cations or halide ions, extensive theoretical studies exist for translational diffusion. On the other hand, theoretical studies exist on rotational diffusion of dipolar molecules with the assumption of point dipole on the centre. The later was generalised by Alavi and Waldeck as already discussed in section II. In this work, we find that the coupling between translational and rotational motion renders interesting features or aspects to the dynamical motion of polyatomic ions. The problem is fairly complex and we have used computer simulations, continuum model and microscopic mode-coupling theory to interrogate various aspects of the problem. The main results of our study are summarised below.

(i) We have employed molecular dynamics simulations to compute the rotational dielectric friction, $\varsigma_{R,DF}$ on three polyatomic ions, namely sulfate, nitrate and acetate ions in water. This study constitutes the first calculation of rotational dielectric friction of polyatomic ions.

(ii) Since the Nee-Zwanzig-Fatuzzo-Mason theories employ a point dipole approximation, we obtained $\varsigma_{R,DF}$ by using the generalized model of Alavi-Waldeck's who derived the rotational dielectric friction using the generalised charge distribution model. We have calculated rotational dielectric friction of the three anions in water and compared the simulation results with the theoretical estimates of Alavi and Waldeck.

(iii) We found that this approach provides poor description for the friction on the sulphate ion but reasonable description for nitrate and acetate ions.

(iv) In order to understand the origin of this failure, we have analysed microscopic salvation structure of the solutes. In our study, we carried out two sets of simulations. In the first set, the real ions with accurate partial charges have been taken and in the second set, we have taken uncharged $SO_4$, $NO_3$ and $CH_3COO$ molecules in water. Finally, we have computed the rotational dielectric friction for all the three ions from simulation by subtracting the total rotational friction in case of uncharged molecule from the total rotational friction for ions. The total rotational friction is obtained from the rotational diffusivities of the solutes.

(v) We have observed that the change in the solvation shell structure for sulphate ion when charges are removed is much more drastic than the mono-valent ions like nitrate, acetate. It is quite easy to understand as because sulfate is a divalent ion having higher charges than nitrate and acetate.

(vi) In order to develop a microscopic theory of rotational friction, we propose a mode-coupling theory formalism to calculate the rotational dielectric friction in a self – consistent scheme. The scheme requires the input of site-site correlation function which can be obtained from RISM. We have outlined the details of the procedure.

(vii) Implementation of MCT requires detailed input of equilibrium two particle correlation functions. These two-particle ion-water correlation functions reveal rich structure.

In a future work, we hope to use the full self-consistent MCT to obtain the total friction on the ions.

## Acknowledgements:

The work was supported partly by Department of Science and Technology (DST), Govt. of India, Sir J. C. Bose fellowship, and Council of Scientific and Industrial Research (CSIR), India.